\documentclass[preprint]{ptephy}%%%%%% to generate preprint number

\preprintnumber{XXXX-XXXX} %%% Insert preprint number here

\usepackage{graphicx}
\usepackage{wrapft}
\def\vc#1{\mbox{\boldmath $#1$}}

\begin{document}

\title{Container structure of $\alpha$$\alpha$$\Lambda$ clusters in ${^9_\Lambda {\rm Be}}$}

\author{\name{Yasuro Funaki}{1}, \name{Taiichi Yamada}{2}, \name{Emiko Hiyama}{1}, \name{Bo Zhou}{3}, and \name{Kiyomi Ikeda}{1}}
%%%%%%%%%%% The \name command should be used as \name{Insert author name here}{Insert affiliation number here}
%%%%% Please use \thanks for contributed author details

%%%%%%%%%%% The \affil command should be used as \affil{Insert affiliation number here}{Insert author address here}
\address{\affil{1}{RIKEN Nishina Center, 2-1 Hirosawa, Wako 351-0198, Japan}
\affil{2}{Laboratory of Physics, Kanto Gakuin University, Yokohama 236-8501, Japan}
\affil{3}{Department of Physics, Nanjing University, Nanjing 210093, China}
\email{funaki@riken.jp}}

\begin{abstract}%
New concept of clustering is discussed in $\Lambda$ hypernuclei using a new-type microscopic cluster model wave function, which has a structure that constituent clusters are confined in a container, whose size is a variational parameter and which we refer to as Hyper-Tohsaki-Horiuchi-Schuck-R\"opke (Hyper-THSR) wave function. 
By using the Hyper-THSR wave function, $2\alpha + \Lambda$ cluster structure in ${^{9}_\Lambda{\rm Be}}$ is investigated. We show that full microscopic solutions in the $2\alpha + \Lambda$ cluster system, which are given as $2\alpha + \Lambda$ Brink-GCM wave functions, are almost perfectly reproduced by the single configurations of the Hyper-THSR wave function. The squared overlaps between the both wave functions are calculated to be $99.5$ \%, $99.4$ \%, and $97.7$ \% for $J^\pi=0^+$, $2^+$, and $4^+$ states, respectively. 
We also simulate the structural change by adding the $\Lambda$ particle, by varying the $\Lambda N$ interaction artificially. As the increase of the $\Lambda N$ interaction, the $\Lambda$ particle gets to move more deeply inside the core and invokes strongly the spatial core shrinkage, and accordingly distinct localized $2\alpha$ clusters appear in the nucleonic intrinsic density, though in ${^{8}{\rm Be}}$ rather gaslike $2\alpha$-cluster structure is shown. The origin of the localization is associated with the strong effect of Pauli principle.
%We conclude that in ${^{9}_\Lambda{\rm Be}}$ the $2\alpha$ clusters move in a container in a nonlocalized way, under the strong influence of inter-$\alpha$ Pauli repulsion, which give rise to an effectively localized clustering in the density distribution. 
We conclude that the container picture of the $2\alpha$ and $\Lambda$ clusters is essential in understanding the cluster structure in ${^{9}_\Lambda{\rm Be}}$, in which the very compact spatial localization of clusters is shown in the density distribution.
\end{abstract}

\subjectindex{xxxx, xxx}

\maketitle

\section{Introduction}\label{intro}

The cluster formation is not only an essential degree of freedom in light nuclei but also remains important in light hypernuclei. While microscopic and semi-microscopic cluster model calculations have a long history in the study of ordinary nuclei~\cite{tang}, they have also played an important role in studying hypernuclei~\cite{HY_PPNP}. In fact, since 1970's, together with the experimental situation, where $(K^-, \pi^-)$ hypernuclear-production reaction data was made available~\cite{bruckner, bertini}, the spectroscopic properties of $p$-shell hypernuclei have been intensively investigated~\cite{PTPS81III} by using the cluster models such as Generator Coordinate Method (GCM)~\cite{GHW} and Orthogonality Condition Model (OCM)~\cite{ocm}. They have revealed novel properties in light hypernuclei. In particular, a ``glue-like'' role of the $\Lambda$ particle and the subsequent core-shrinkage effect have been widely discussed in ${^7_\Lambda{\rm He}}$, ${^7_\Lambda{\rm Li}}$, ${^7_\Lambda{\rm Be}}$, ${^9_\Lambda{\rm Be}}$, ${^{13}_\Lambda{\rm C}}$, ${^{20}_\Lambda{\rm Ne}}$, ${^{21}_\Lambda{\rm Ne}}$, etc~\cite{YMB_PTP76, MBI_PTP, sakuda, PTP71,PTPS81IV, hiyama_PRC53, hiyama_PRC59, hiyama_PRL}. The reduction of $B(E2)$ value, due to the effect, was predicted theoretically~\cite{MBI_PTP, hiyama_PRC59}, which was later on confirmed by experiment in ${^7_\Lambda{\rm Li}}$~\cite{tanida}. As a typical example of cluster states, many authors investigated ${^9_\Lambda{\rm Be}}$ hypernucleus by using the GCM and OCM~\cite{PTP66,PTP69,PTP73,MBI_PTP,YMB_PTP76,hiyama97}. They showed the rotational spectrum composed of the $2\alpha$ clusters, in an analogy to the case of ${^8{\rm Be}}$, and the advent of unique state called the genuinely hypernuclear state~\cite{PTP69} or the supersymmetric state~\cite{gal}.
%They showed the rotational spectrum composed of the $2\alpha$ clusters, in an analogy to the case of ${^8{\rm Be}}$, and the advent of unique state called the genuinely hypernuclear state, which was originally pointed out by Dalitz and Gal~\cite{gal}.
 %The appearance of the genuinely hypernuclear states was also reported in ${^{13}_\Lambda{\rm C}}$ and ${^{21}_\Lambda{\rm Ne}}$~\cite{PTP69}. 
The importance of the clustering aspect in ${^9_\Lambda{\rm Be}}$ was also confirmed quantitatively in comparison with the experimental data of $(K^-,\pi^-)$, $(\pi^+,K^+)$, and $(K^{-}\ {\rm stopped},\pi^-)$ reactions~\cite{PTP73}.

On the other hand, in ordinary light nuclei, a very promising model wave function is proposed to describe cluster states. That is the so-called Tohsaki-Horiuchi-Schuck-R\"opke (THSR) wave function~\cite{thsr}, which was originally introduced to investigate the $\alpha$ condensate states with a dilute gaslike configuration of a few number of $\alpha$ particles~\cite{yamal,YS_epja,funaki_8Be,funaki_12C,funaki_concept}. The wave function is now known to be the best wave function to describe such the special cluster structure realized in ${^8{\rm Be}}$ and the Hoyle state (the second $0^+$ state at $7.65$ MeV of ${^{12}{\rm C}}$). For both cases the Resonating Group Method (RGM) wave function~\cite{rgm,kamimura}, which was obtained in the full model space with respect to the $\alpha$-$\alpha$ relative motions, is almost $100$ \% equivalent to a single and energetically optimal configuration of the THSR wave function with the condensate character~\cite{funaki_8Be,funaki_12C,funaki_concept}. 

On the contrary, non-gaslike cluster states have been recognized as having localized structures of clusters, which are obviously quite different from the $\alpha$ condensate states mentioned above.
% mentioned nonlocalized structure with gaslike cluster configuration. 
The inversion doublet band states of $\alpha + {^{16}{\rm O}}$ in ${^{20}{\rm Ne}}$~\cite{HI_20Ne}, linear-chain structures of $3\alpha$ and $4\alpha$ particles~\cite{morinaga}, etc. are such typical examples. Nevertheless recent calculations clarified that all those structure states can also be expressed by the single configurations of the THSR wave function with nearly $100$ \% accuracy~\cite{zhou_bo1,zhou_bo2,suhara}. This is amazing since the THSR wave function has always been considered to provide a nonlocalization picture based on a container structure by its functional form and to be specialized in describing the gaslike cluster states. These results therefore urge us to reexamine our conventional picture about the localized cluster structures.
%, which has been considered to be specialized in the gaslike states like the $\alpha$ condensates. 
It is then interesting to see how much the container picture works well for cluster states in hypernuclei, since additional $\Lambda$ particle is known to make spatial core shrinkage and thus seems to realize more localized cluster structure. 

The purpose of this study is to introduce a THSR-type model wave function in $\Lambda$ hypernuclei, as the Hyper-THSR wave function, and to apply it to ${^9_\Lambda{\rm Be}}$. The effect of spatial core shrinkage, which will be invoked by the additional $\Lambda$ particle, is expected to be properly taken into account by the Hyper-THSR wave function, since the only parameter in the wave function is what specifies monopole-like dilatation of a whole nucleus. We first perform the full microscopic $\alpha +\alpha+\Lambda$ cluster model calculation based on the $\alpha + \alpha + \Lambda$ Brink model wave function and GCM~\cite{brink,margenau}. We then compare the full solution with the single Hyper-THSR wave function and discuss the container structure of the $2\alpha$ and $\Lambda$ clusters. We also discuss the intrinsic density of ${^9_\Lambda{\rm Be}}$ and demonstrate the structural change from ${^8{\rm Be}}$ to ${^9_\Lambda{\rm Be}}$, in which the roles of $\Lambda$ particle and Pauli principle to give the spatial localization of the $2\alpha$ clusters is emphasized. This paper is organized as follows: In \S~\ref{Hyp_THSR} we extend the deformed THSR wave function to the Hyper-THSR wave function so as to apply to $n\alpha + \Lambda$ system. We then outline the $\alpha + \alpha + \Lambda$ Brink wave function and GCM in \S~\ref{brink_gcm}, and microscopic Hamiltonian we adopt, in \S~\ref{haml}. In \S~\ref{result}, we compare the $\alpha + \alpha + \Lambda$ Hyper-THSR wave function with the Brink-GCM wave function. We also discuss the effect of the $\Lambda$ particle injected to the ${^8{\rm Be}}$ core, and the change of the intrinsic structures. \S~\ref{summary} is devoted to Summary.

%%%%%%%%%%%%%%%%   Section 2  %%%%%%%%%%%%%%%%%%%
\section{Formulation}

\subsection{Hyper-THSR wave function}\label{Hyp_THSR}

%%%%%%%%%%%%%%%%%%%%%%%%%%%%%%%%%%%%%%%%%%%%%%%%%%

 We propose a new-type microscopic cluster model wave function, which we refer to as the Hyper-THSR wave function. This is based on the deformed $n\alpha$ THSR wave function in a $4n$-nucleus~\cite{thsr,funaki_8Be}, which is characterized by two kinds of size parameters, one for the $\alpha$ particle and the other for the center-of-mass motions of the $\alpha$ particles. The former is denoted as $b$ and the latter as $B_k$ $(k=x,y,z)$, with deformation being taken into account. The explicit form of the THSR wave function is shown as follows:
\begin{equation}
 \Phi_{n\alpha}^{\rm THSR} \propto {\cal A}\ \Big\{ \prod_{i=1}^n\exp \Big[- 
\sum_{k=x,y,z}\frac{2}{B_k^2} (X_{ik}-X_{Gk})^2 \Big] \phi(\alpha_i) \Big\}, 
 \label{eq:thsr} 
\end{equation}
and $\phi(\alpha_i)$ is the internal wave function of the $i$-th $\alpha$ particle, with $(0s)^4$ shell-model configuration,
\begin{equation}
\phi(\alpha_i) \propto \exp\Big[-\sum_{1\leq k<l \leq4}({\vc r}_{i,k} - 
{\vc r}_{i,l})^2/(8b^2)\Big]. \label{eq:int_alpha}
\end{equation}
Here $\vc{X}_i$ denotes the center-of-mass coordinate of the $i$-th $\alpha$ particle, and the spurious total center-of-mass motion, which is concerned with the center-of-mass coordinate $\vc{X}_G$, is exactly eliminated in Eq.~(\ref{eq:thsr}).
%In Eq.~(\ref{eq:int_alpha}) the two protons and two neutrons in the $\alpha$ particle sit in an $S$-wave with the size parameter $b$, which is fixed at the size in free space. 
The operator ${\cal A}$ antisymmetrizes all the $4n$ nucleons, and therefore, when $B_x=B_y=B_z \rightarrow b$, the normalized THSR wave function coincides with the shell model Slater determinant. On the contrary, so far as $|\vc{B}|(=(B_x^2+B_y^2+B_z^2)^{1/2})$ is large enough to be able to neglect the effect of the antisymmetrizer ${\cal A}$, all $\alpha$ clusters occupy an identical deformed orbit, $\exp [-\sum_{k=x,y,z}$ $\frac{2}{B_k^2}(X_k-X_{Gk})^2 ]$, with respect to the $\alpha$ particle's center-of-mass motions measured from the total center-of-mass position. This displays a product arrangement of the $n \alpha$ particles, and hence is referred to as the $n\alpha$ condensed state~\cite{funaki_concept}.
In all the subsequent calculations, however, from a technical reason, we parametrize the THSR wave function by $\beta_k$ $(k=x,y,z)$ and $b$, instead of $B_k$ and $b$, following the relation $B_k^2=b^2+2\beta_k^2$ $(k=x,y,z)$. Of course, this does not change any picture of this model wave function mentioned above.

We should repeat that the THSR wave function provides a structure that the $\alpha$ clusters are confined in a container, whose size is characterized by the variational parameter $\vc{\beta}$, in a nonlocalized way, and occupy an identical orbit of a self-consistent mean-field potential of the clusters, under the effect of the antisymmetrization. This is very different from the conventional cluster model wave function, like the Brink model wave function~\cite{brink}, in which relative motions of clusters are characterized by inter-cluster distance parameters, in a localized way. As mentioned in \S \ref{intro}, the THSR wave function, however, very nicely describes not only the loosely bound $\alpha$ cluster states such as ${^{8}{\rm Be}}$ and the Hoyle state, but also rather compact cluster states like the $\alpha+{^{16}{\rm O}}$ inversion doublet band states and $\alpha$-linear-chain states with practically $100$ \% accuracy, although the latters had been considered to have the localized cluster structures. Even the ground state of ${^{12}{\rm C}}$ is also shown to be described very precisely by the single THSR wave function with a proper choice of $\vc{\beta}$ value~\cite{funaki_concept}. All these imply that the parameter $\vc{\beta}$ plays a role of dynamical coordinate specifying a monopole-like dilatation of whole system, to describe compact cluster states to dilute cluster states in a unified way.

%The wave function Eq.~(\ref{eq:thsr}) succeeded in describing gaslike cluster states realized in light $N=Z$ nuclei such as ${^{8}{\rm Be}}$, ${^{12}{\rm C}}$, ${^{16}{\rm O}}$ and ${^{20}{\rm Ne}}$~\cite{funaki_concept,4athsr}. In particular the first excited $0^+$ state in ${^{12}{\rm C}}$ is described correctly by this wave function, with a loosely coupled structure of the three $\alpha$ particles reminiscent of a gas occupying the lowest $0S$ orbit of a mean-field potential for the $\alpha$ particle~\cite{funaki_12C}. Apart from this example, the THSR wave function is known to give a very good description of the ground states of those nuclei since their shell model configurations are properly taken into account due to the antisymmetrization of nucleons, together with the $\alpha$-like ground state correlations. In particular, for the ground state rotational bands of ${^{12}{\rm C}}$~\cite{funaki_concept} and ${^{20}{\rm Ne}}$~\cite{zhou_bo1,zhou_bo2}, the THSR wave functions give large squared overlap (close to 100 \%) with the corresponding microscopic cluster model wave functions such as RGM (Resonating Group Method) and GCM (Generator Coordinate Method)~\cite{carbon}.

The Hyper-THSR wave function describing the $2\alpha+\Lambda$ hypernucleus with good angular momentum is then introduced as follows:
\begin{eqnarray}
&&\Psi_J^{\rm H} (\vc{\beta}) = {\hat P}^{J}_{MK} \Phi_{2\alpha-\Lambda}^{\rm H-THSR}(\vc{\beta}), \nonumber \\
&&\Phi_{2\alpha-\Lambda}^{\rm H-THSR} (\vc{\beta}) =  \Phi_{2\alpha}^{\rm THSR}(\vc{\beta}) \sum_{\kappa}f_{\Lambda}(\vc{\beta}, \kappa)\varphi_\Lambda(\kappa), \label{eq:hypthsr} 
\end{eqnarray}
where ${\hat P}^J_{MK}$ is the angular-momentum projection operator and the $\Lambda$ particle simply couples to the $2\alpha$ core nucleus in an $S$ wave. Its radial part is expanded in terms of Gaussian basis functions, $\varphi_\Lambda(\kappa)=(\pi/2\kappa)^{-3/4} \exp(-\kappa r_{2\alpha-\Lambda}^2)$ with respect to the width parameter $\kappa$, where $\vc{r}_{2\alpha-\Lambda}=\vc{r}_\Lambda - (\vc{X}_1+\vc{X}_2)/2$. In the practical calculations, we assume an axially-symmetric deformation in Eq.~(\ref{eq:thsr}), $\beta_x=\beta_y\equiv \beta_\perp \ne \beta_z$ and fix the parameter $b=1.36$ fm in Eq.~(\ref{eq:int_alpha}), which is the same value as adopted in Ref.~\cite{funaki_8Be} and almost the same as the size of the $\alpha$ particle in free space. Note that the Hyper-THSR wave function, Eq.~(\ref{eq:hypthsr}), has positive intrinsic parity. The coefficients of the expansion $f_{\Lambda}(\beta_\perp,\beta_z, \kappa)$ are then determined by solving the following Hill-Wheeler-type equation of motion~\cite{GHW}, 
\begin{eqnarray}
&&\sum_{\kappa^\prime} \big\langle {\hat P}^J_{M0}\Phi_{2\alpha}^{\rm THSR} (\beta_\perp,\beta_z)\varphi_\Lambda(\kappa) \big|{\hat H}-E(\beta_\perp,\beta_z) \big|{\hat P}^J_{M0} \Phi_{2\alpha}^{\rm THSR} (\beta_\perp,\beta_z)\varphi_\Lambda(\kappa^\prime) \big\rangle  \nonumber \\ 
&& \hspace{8cm} \times f_{\Lambda}(\beta_\perp,\beta_z,\kappa^\prime)=0. \label{eq:ghweq}
\end{eqnarray}
This Hyper-THSR wave function is characterized only by the parameters $\beta_\perp$ and $\beta_z$, which correspond to a spatial extension of the whole nucleus. It is reported that the $\Lambda$ particle invokes spatial core shrinkage in many hypernuclei. It should be mentioned that such a core shrinkage effect is expected to be taken into account very naturally by this parametrization of $\vc{\beta}$, in general hypernuclei, since it specifies the dilatation of the whole nucleus. In the present work, only the $S$-wave component of the $\Lambda$ particle is considered, for simplicity. Thus the monopole-like shrinkage is expected to be described nicely by this wave function. The extension to inclusion of the other angular-momentum channels is of course possible and left as a future work. 

\subsection{Brink-GCM wave function}\label{brink_gcm}
We briefly mention here the conventional microscopic $2\alpha+\Lambda$ cluster model, which is Brink-GCM wave function and is used to compare with the Hyper-THSR wave function in the subsequent sections. The $2\alpha + \Lambda$ Brink-GCM wave function is based on the following Brink cluster model wave functions,
\begin{equation}
u_L(\vc{R})={\widehat P}^L_{M0} {\cal A}\Big[\exp \Big\{-\frac{(\vc{X}-\vc{R})^2}{b^2} \Big\} \phi(\alpha_1)\phi(\alpha_2) \Big],
\end{equation}
for the $2\alpha$ clusters and 
\begin{equation}
\psi_\lambda(\vc{S})={\widehat P}^\lambda_{\mu \nu} \exp \Big\{-\frac{4\rho}{b^2(8+\rho)}(\vc{r}_{2\alpha-\Lambda}-\vc{S})^2\Big\},
\end{equation}
for the $\Lambda$ particle, where the Jacobi coordinates and a mass ratio between nucleon $(M_N)$ and $\Lambda$-particle $(M_\Lambda)$ are defined by $\vc{X}=\vc{X}_1-\vc{X}_2$ and $\vc{r}_{2\alpha-\Lambda}=\vc{r}_\Lambda - (\vc{X}_1+\vc{X}_2)/2$, and $\rho=M_\Lambda/M_N$, respectively. These Brink wave functions elucidate the localized clustering, where the $\alpha$ clusters and $\Lambda$ particle are localized around a separation distance $R$ for $\alpha$-$\alpha$ part and $S$ for $\Lambda$-${^8{\rm Be}}$ part, respectively, with a rather small width parameter $b(=1.36\ {\rm fm})$. The Brink-GCM wave function can then be described by a superposition of these localized cluster wave functions, with discretized values for the radial parts of the separation-distance parameters $R$ and $S$, as follows:
\begin{equation}
\Psi_J^{\rm B}= \sum_{L,\lambda} \sum_{R,S} f^{(L,\lambda)}(R,S) [u_L(\vc{R}), \psi_\lambda(\vc{S})]_J. \label{eq:b-gcm_wf}
\end{equation}
The coefficients $f^{(L,\lambda)}(R,S)$ in Eq.~(\ref{eq:b-gcm_wf}) are determined by solving the following Hill-Wheeler equation,
\begin{equation}
\sum_{R^\prime,S^\prime}\sum_{L^\prime,\lambda^\prime} \big\langle [u_L(\vc{R}),\psi_\lambda(\vc{S}))]_J  \big|{\hat H}-E \big| [u_{L^\prime}(\vc{R}^\prime),\psi_{\lambda^\prime}(\vc{S}^\prime))]_J  \big\rangle f^{(L^\prime,\lambda^\prime)}(R^\prime,S^\prime)=0. \label{eq:b-gcm}
\end{equation}
The weight of contribution from an angular-momentum channel $(L,\lambda)$ in the Brink-GCM wave function $\Psi_J^{\rm B}$ of Eq.~(\ref{eq:b-gcm_wf}) can be defined as follows:
\begin{equation}
w^2_{L,\lambda}=\sum_{R,S}  \big\langle [u_L(\vc{R}), \psi_\lambda(\vc{S})]_J \big| [u_L(\vc{R}), \psi_\lambda(\vc{S})]_J \big\rangle \big|f^{(L,\lambda)}(R,S) \big|^2. \label{eq:w2}
\end{equation}

\subsection{Microscopic Hamiltonian}\label{haml}
We use the following microscopic Hamiltonian for ${^{9}_\Lambda{\rm Be}}$, composed of kinetic energies $-\frac{\hbar^2}{2M_N} \nabla_i^2$ and $-\frac{\hbar^2}{2M_\Lambda}\nabla_\Lambda^2$, the effective nucleon-nucleon interaction $V^{(NN)}_{ij}$, the Coulomb potential $V_{ij}^{(C)}$, and the $\Lambda N$ interaction $V^{(\Lambda N)}_{i}$:
\begin{equation}
H=-\sum_{i=1}^{8}\frac{\hbar^2}{2M_N}\nabla_i^2 -\frac{\hbar^2}{2M_\Lambda}\nabla_\Lambda^2 
- T_G +\sum_{i<j}^{8}V_{ij}^{(C)} + \sum_{i<j}^{8} V^{(NN)}_{ij} + \sum_{i=1}^{8}V_{i}^{(\Lambda N)}, \label{eq:hml} 
\end{equation}
where the center-of-mass kinetic energy $T_G$ is subtracted. We neglect the small $\Lambda N$ spin-orbit interaction. For the $NN$ interaction, we adopt the Volkov No. 1 force~\cite{volkov} with the Majorana parameter value $M=0.56$, which is the same as used in the previous study of ${^{8}{\rm Be}}$~\cite{funaki_8Be}. For the $\Lambda N$ interaction, we adopt two kinds of spin-independent parts of the YNG interactions, Nijmegen model-D (ND) and J\"ulich version-A (JA)~\cite{yamamoto}, since they are well tested in structural calculations~\cite{HY_PPNP,hiyama97,yamamoto,isaka} and it is shown that the use of ND and JA gives the deepest and shallowest $\Lambda$ binding energies of the versions adopted in Ref.~\cite{hiyama97}, respectively.
%The use of JA is also shown to reproduce well the $\Lambda$ binding energy in light hypernuclei~\cite{hiyama97}.}
The fermi-momentum parameter $k_F$, which appears in the YNG interactions, is determined by $\alpha + \Lambda$ cluster model, so as to reproduce the empirical value of $\Lambda$ binding energy of ${^5_\Lambda {\rm He}}$, i.e. $B_\Lambda({^5_\Lambda{\rm He}})=3.12$ MeV. The adopted values are $k_F=0.962$ ${\rm fm}^{-1}$ for ND and $k_F=0.757$ ${\rm fm}^{-1}$ for JA. 

\section{Results and Discussion}\label{result}

\subsection{Results of Brink-GCM calculation}

Following Eq.~(\ref{eq:b-gcm}), we perform the GCM calculations with the bases of the Brink wave function Eq.~(\ref{eq:b-gcm_wf}), for $J^\pi=0^+$, $2^+$, and $4^+$ states. We adopt $3$, $6$, and $6$ sets of angular-momentum channels $(L,\lambda)$, for the $\alpha$-$\alpha$ $(L)$ and ${^{8}{\rm Be}}$-$\Lambda$ $(\lambda)$ relative motions, for the $0^+$, $2^+$, and $4^+$ states, respectively. We also adopt the radial parts of the generator coordinates, $R= i$ fm with $i=1 - 10$ and $S=0.5 + i-1$ fm with $i=1 - 7$ for the $0^+$ and $2^+$ states. For the resonant $4^+$ state, $R=i$ fm with $i=1 - 5$ and $S=0.5 + i-1$ fm with $i=1 - 5$ are adopted so that admixture of spurious scattering components can be avoided within the bound state approximation. Energies are converged to within tens of keV for all those states. 

In Table~\ref{tab1}, we show the calculated binding energies for both cases of the adopted $\Lambda N$ interactions ND and JA. Throughout the $0^+$, $2^+$, and $4^+$ states, the choice of potential ND gives deeper binding energies than that of JA by about $0.8 \sim 0.9$ MeV. The potential JA gives much closer $\Lambda$ binding energy of the ground state to the experimental value, $B_\Lambda({^9_\Lambda{\rm Be}})=6.71$ MeV, than the potential ND. This is the same situation as in the previous results studied with the $\alpha+\alpha+\Lambda$ OCM~\cite{hiyama97}. We list in Table~\ref{tab1} the weight $w^2_{L, \lambda}$ in Eq.~(\ref{eq:w2}). We can see that for all the $J^+$ states the total wave functions are dominated practically only by the $S$-wave channel for the $\Lambda$ particle, i.e. $(L, \lambda)=(J, 0)$. In particular, for the choice of potential ND, the $S$-wave dominates by more than $96$ \%. This is consistent with the previous results obtained by the GCM calculation with the $\alpha+\alpha+\Lambda$ microscopic cluster model, in which a phenomenological $\Lambda N$ potential with one-range Gaussian form is used~\cite{PTP66}.

\subsection{Energy surfaces for Hyper-THSR wave function}

\begin{table}[htbp]
\begin{center}
\caption{The weight $w_{L,\lambda}^2$ of a channel specified by the angular momenta of $\alpha$-$\alpha$ part $(L)$ and $\Lambda$-${^8{\rm Be}}$ part $(\lambda)$, defined by Eq.~(\ref{eq:w2}). The total binding energies $E$ and $\Lambda$ binding energy $B_\Lambda$, given by the Brink-GCM wave function, are also shown. $B_\Lambda$ is defined as the binding energy measured from the calculated ${^8{\rm Be}}$ energy $-54.45$ MeV. Energies in parentheses are the results of the single channel calculation of $(L,\lambda)=(J,0)$. The two kinds of the $\Lambda N$ interaction, YNG-ND and -JA are used.} \label{tab1}
\begin{tabular}{ccccccccccc}
\hline\hline
$J^\pi$ &  & $E$ & $B_\Lambda$ &  & \multicolumn{6}{c}{$w^2_{L, \lambda}$} \\
\hline
 &  &  &  & $(L, \lambda)$ & \multicolumn{2}{c}{$(0, 0)$} & \multicolumn{2}{c}{$(2, 2)$} & \multicolumn{2}{c}{$(4, 4)$} \\
 & \raisebox{-1.8ex}[0pt][0pt]{ND} & $-61.78$ & $7.33$ &  & \multicolumn{2}{c}{\raisebox{-1.8ex}[0pt][0pt]{$0.973$}} & \multicolumn{2}{c}{\raisebox{-1.8ex}[0pt][0pt]{$0.026$}} & \multicolumn{2}{c}{\raisebox{-1.8ex}[0pt][0pt]{$4$$\times$$10^{-4}$}} \\
$0^+$ &  & $(-61.14)$ &  &  &  &  &  &  &  &  \\
 & \raisebox{-1.8ex}[0pt][0pt]{JA} & $-60.98$ & $6.53$ &  & \multicolumn{2}{c}{\raisebox{-1.8ex}[0pt][0pt]{$0.951$}} & \multicolumn{2}{c}{\raisebox{-1.8ex}[0pt][0pt]{$0.048$}} & \multicolumn{2}{c}{\raisebox{-1.8ex}[0pt][0pt]{$9$$\times$$10^{-4}$}} \\
 &  & $(-59.78)$ &  &  &  &  &  &  &  &  \\
\hline
 &  &  &  & $(L, \lambda)$ & $(2, 0)$ & $(0, 2)$ & $(2, 2)$ & $(2, 4)$ & $(4, 2)$ & $(4, 4)$ \\
 & \raisebox{-1.8ex}[0pt][0pt]{ND} & $-58.90$ & $4.45$ &  & \raisebox{-1.8ex}[0pt][0pt]{$0.968$} & \raisebox{-1.8ex}[0pt][0pt]{$0.012$} & \raisebox{-1.8ex}[0pt][0pt]{$0.011$} & \raisebox{-1.8ex}[0pt][0pt]{$3$$\times$$10^{-4}$} & \raisebox{-1.8ex}[0pt][0pt]{$0.009$} & \raisebox{-1.8ex}[0pt][0pt]{$2$$\times$$10^{-4}$} \\
$2^+$ &  & $(-58.25)$ &  &  &  &  &  &  &  &  \\
 & \raisebox{-1.8ex}[0pt][0pt]{JA} & $-58.10$ & $3.65$ &  & \raisebox{-1.8ex}[0pt][0pt]{$0.941$} & \raisebox{-1.8ex}[0pt][0pt]{$0.022$} & \raisebox{-1.8ex}[0pt][0pt]{$0.021$} & \raisebox{-1.8ex}[0pt][0pt]{$7$$\times$$10^{-4}$} & \raisebox{-1.8ex}[0pt][0pt]{$0.016$} & \raisebox{-1.8ex}[0pt][0pt]{$4$$\times$$10^{-4}$} \\
 &  & $(-56.87)$ &  &  &  &  &  &  &  &  \\
\hline
 &  &  &  & $(L, \lambda)$ & $(4, 0)$ & $(0, 4)$ & $(2, 2)$ & $(2, 4)$ & $(4, 2)$ & $(4, 4)$ \\
 & \raisebox{-1.8ex}[0pt][0pt]{ND} & $-51.47$ & $-2.98$ &  & \raisebox{-1.8ex}[0pt][0pt]{$0.965$} & \raisebox{-1.8ex}[0pt][0pt]{$2$$\times$$10^{-4}$} & \raisebox{-1.8ex}[0pt][0pt]{$0.027$} & \raisebox{-1.8ex}[0pt][0pt]{$2$$\times$$10^{-4}$} & \raisebox{-1.8ex}[0pt][0pt]{$0.007$} & \raisebox{-1.8ex}[0pt][0pt]{$9$$\times$$10^{-4}$} \\
$4^+$ &  & $(-51.00)$ &  &  &  &  &  &  &  &  \\
 & \raisebox{-1.8ex}[0pt][0pt]{JA} & $-50.50$ & $-3.95$ &  & \raisebox{-1.8ex}[0pt][0pt]{$0.936$} & \raisebox{-1.8ex}[0pt][0pt]{$4$$\times$$10^{-4}$} & \raisebox{-1.8ex}[0pt][0pt]{$0.049$} & \raisebox{-1.8ex}[0pt][0pt]{$4$$\times$$10^{-4}$} & \raisebox{-1.8ex}[0pt][0pt]{$0.014$} & \raisebox{-1.8ex}[0pt][0pt]{$2$$\times$$10^{-4}$} \\
 &  & $(-49.52)$ &  &  &  &  &  &  &  &  \\
\hline\hline
\end{tabular}
\end{center}
\end{table}

\begin{figure}[htbp]
\begin{center}
\includegraphics[scale=0.70]{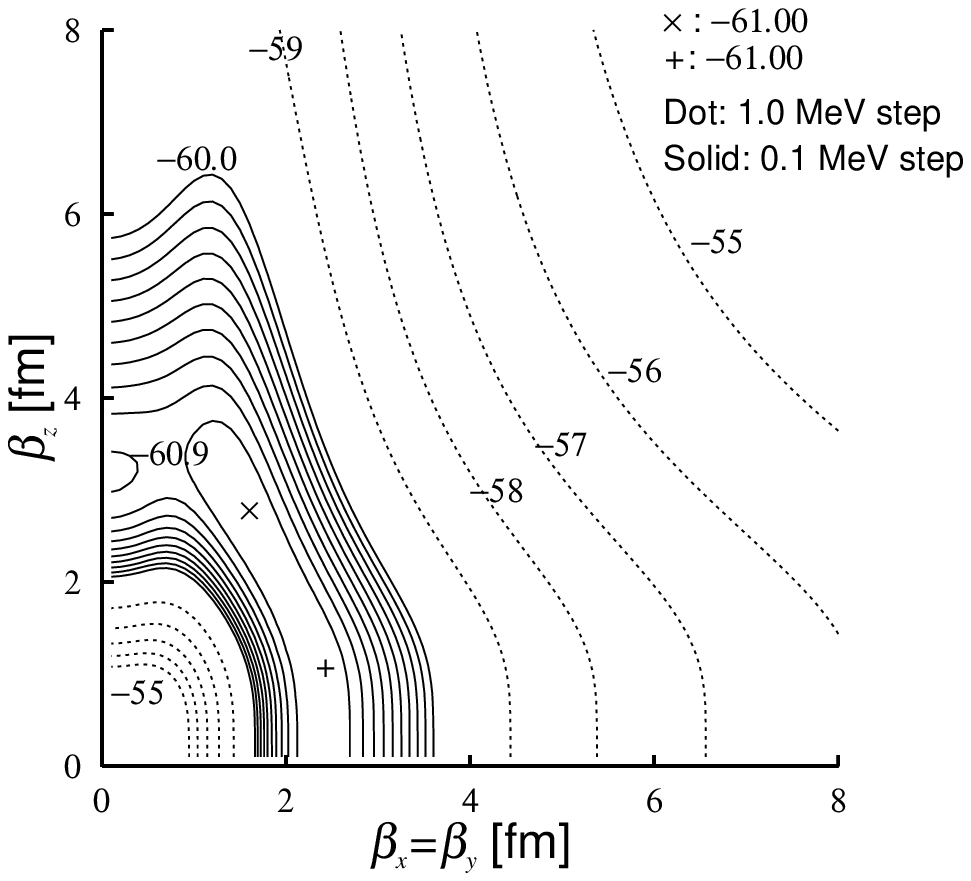}
\label{fig:energy1}
\includegraphics[scale=0.70]{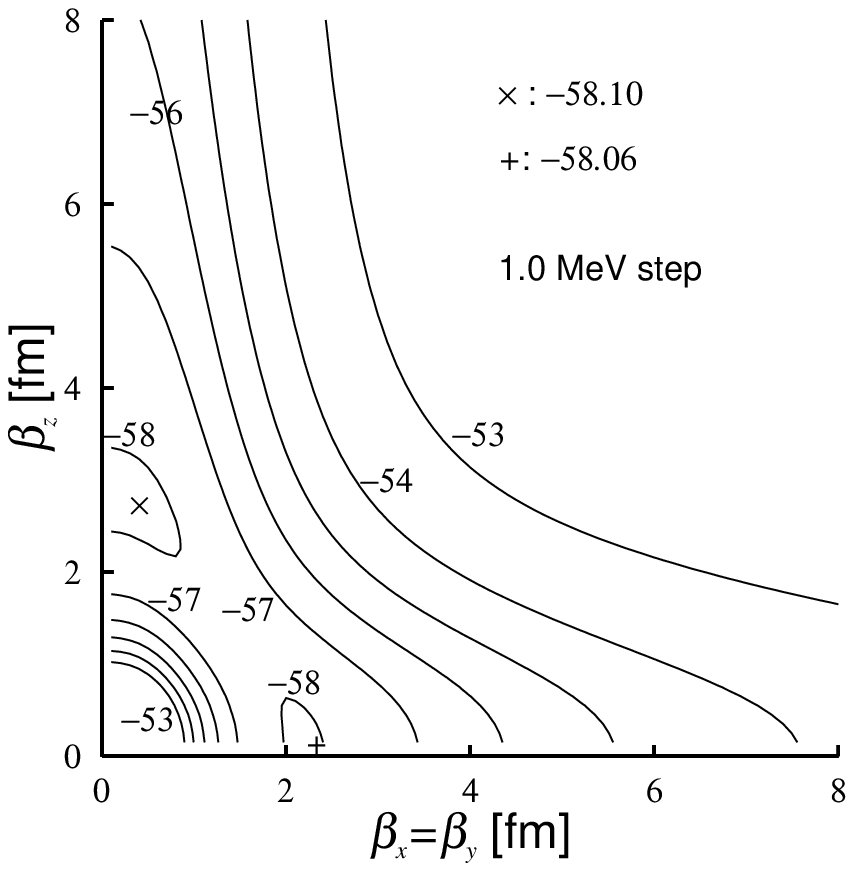}
\label{fig:energy2}
\includegraphics[scale=0.70]{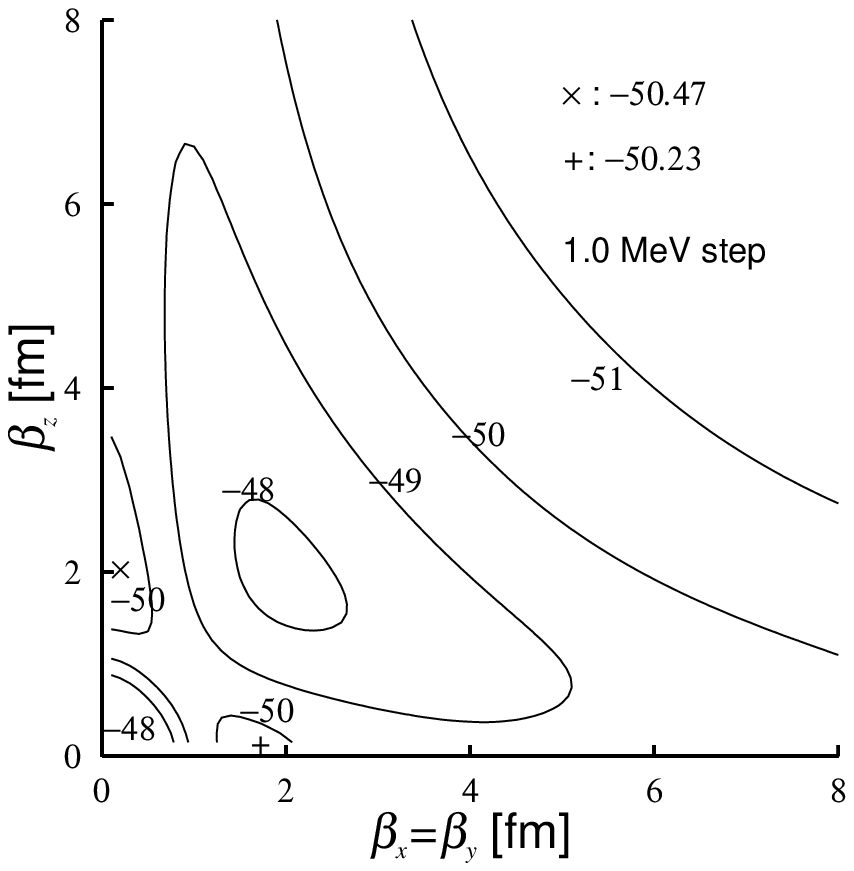}
\label{fig:energy3}
\caption{Contour maps of the energy surfaces for $J^\pi=0^+$(left top), $2^+$(right top), and $4^+$(bottom) states in the two-parameter space $\beta_x=\beta_y(\equiv \beta_\perp), \beta_z$, given as $E(\beta_\perp,\beta_z)$ in the calculation of Eq.~(\ref{eq:ghweq}). Two minima are denoted by $\times$ and $+$. YNG-ND is adopted for the $\Lambda N$ interaction.}\label{fig:energy}
\end{center}
\end{figure}
In order to compare the single Hyper-THSR wave function with the Brink-GCM wave functions obtained above, we first solve the Hill-Wheeler-type equation of motion Eq.~(\ref{eq:ghweq}). We show the contour map of the lowest eigenvalues $E(\beta_\perp, \beta_z)$ in the two-parameter space $\beta_\perp$ and $\beta_z$, in Fig.~\ref{fig:energy}, for $J^\pi=0^+$ (left top), $2^+$ (right top), and $4^+$ (bottom) states. We take the following discretized values for $\kappa$ in solving Eq.~(\ref{eq:ghweq}): $\kappa^{-1/2}=1.9\times 1.25^{n-1}$ fm for $n=1,\cdots, 5$. 
%Here the eigenfunction $\Psi_J^H(\beta_\perp, \beta_z)$ is the single Hyper-THSR wave function defined in Eq.~(\ref{eq:hypthsr}). 
We can see that for all the $J^+$ states, two minima (denoted as $\times$ and $+$) appear in the prolately deformed region of $\beta_z > \beta_\perp$ and the oblately deformed region of $\beta_z < \beta_\perp$. The minimum energies and the $\vc{\beta}$ values giving them are also listed in Table~\ref{tab2}. The region which connects the two minima is energetically flat within several hundred keV. This, however, does not mean that a different excited state appears in the same energy region but is a peculiar feature of the THSR-type wave function. In Ref.~\cite{funaki_8Be} it was shown for ${^8{\rm Be}}$ that angular-momentum-projected THSR wave functions from prolate deformation are practically identical to the ones from oblate deformation. In fact, the mutual squared overlaps between the both wave functions giving the minima were then calculated to be more than $99$ \%.
It was further argued recently that in two-cluster systems the THSR wave function with oblate intrinsic deformation can be generated from rotation of the THSR wave function with intrinsic prolate deformation with respect to an axis which is perpendicular to a symmetry axis~\cite{zhou_bo3}. We can thus say that the intrinsic shape of ${^{9}_\Lambda {\rm Be}}$ is not oblate but prolate deformation with $2\alpha + \Lambda$ structure, as naturally expected in two-cluster systems. We will later discuss the intrinsic shape of ${^9_\Lambda {\rm Be}}$.

We show the energies obtained by the single channel calculations of $(L,\lambda)=(J,0)$ via Brink-GCM ansatz in parentheses of Table~\ref{tab1}. Compared with them, the minimum values in Fig.~\ref{fig:energy} are only slightly higher, by less than about $0.2$ MeV for the $0^+$ and $2^+$ states and $0.5$ MeV for the $4^+$ state. Considering the fact that the minimum energies are expressed by only one configuration of the Hyper-THSR wave function, we can say that the Hyper-THSR ansatz works very well and nicely reproduces the $S$-wave sectors of the Brink-GCM calculations.

%Considering the fact that the $S$-wave components of the $\Lambda$-particle are dominant for these states but still several percent of mixture from other angular-momentum channels remains, the lack of the energy for the single Hyper-THSR wave functions is reasonable and we can say that they nicely reproduce the $S$-wave sectors of the Brink-GCM wave functions.

While these contour maps of energy surfaces resemble those of ${^8{\rm Be}}$ which are shown in Ref.~\cite{funaki_8Be}, the values $\vc{\beta}$ giving the minima shown in Table~\ref{tab2} are very different from those in ${^8{\rm Be}}$. In the case of $J^\pi=0^+$ state, the former is $\beta_\perp=1.5$ fm and $\beta_z=2.8$ fm for the choice of potential ND and the latter is $\beta_\perp=1.8$ fm and $\beta_z=7.8$ fm. This smaller value of $\vc{\beta}$ is of course because of the injected $\Lambda$ particle, which shrinks the core nucleus ${^8{\rm Be}}$. This shrinkage can also be seen from the calculated rms radius of the core $R^{\rm (c)}_{\rm rms}=2.31$ fm, which is much smaller than that of ${^8{\rm Be}}$, $R_{\rm rms}=2.87$ fm, for the ground state. 

The values of $\vc{\beta}$ giving the minima do not depend on the choice of $\Lambda N$ potential, but they take smaller values as the increase of $J$. The $J^\pi=0^+$ state gives rather large $\beta_\perp$ value at the minimum in prolately deformed region, while the $J^\pi=2^+,4^+$ states give values of $\beta_\perp \sim 0$ and slightly smaller $\beta_z$ value than that for the $J^\pi=0^+$ state. Accordingly the rms radii of the core $R^{\rm (c)}_{\rm rms}$ in Table~\ref{tab2} get slightly smaller as the increase of $J$. We can also see that the rms distances between the ${^8{\rm Be}}$ core and $\Lambda$ particle, $R^{({\rm c}-\Lambda)}_{\rm rms}$, get slightly smaller as the increase of $J$, so as to gain the binding energies by making larger the overlap between the core and $\Lambda$ particle.

\subsection{Squared overlaps between Hyper-THSR and Brink-GCM wave functions}

\begin{table}[htbp]
\begin{center}
\caption{The minimum binding energies of $E(\beta_\perp,\beta_z)$, the corresponding $\Lambda$ binding energies $B_\Lambda$, the rms radii of the ${^8{\rm Be}}$ core $R^{\rm (c)}_{\rm rms}$, the rms distances between the core and $\Lambda$ particle $R^{({\rm c}-\Lambda)}_{\rm rms}$, and the rms radii of ${^9_\Lambda{\rm Be}}$ $R_{\rm rms}$ at the minimum positions are shown, together with the corresponding $\beta_\perp$ and $\beta_z$ values. The maximum squared overlap values of ${\cal O}_J(\beta_\perp,\beta_z)$ defined by Eq.~(\ref{eq:ovlp}), are also shown, together with the $\beta_\perp$ and $\beta_z$ values giving the maxima. The two kinds of the $\Lambda N$ interaction, YNG-ND and -JA are adopted.}\label{tab2}
\begin{tabular}{ccclcccccl}
\hline\hline
$J^\pi$ &  & \multicolumn{2}{c}{$E$ $(\beta_\perp, \beta_z)$} & \multicolumn{1}{c}{$B_\Lambda$} & $R^{\rm (c)}_{\rm rms}$ & $R^{({\rm c}-\Lambda)}_{\rm rms}$ & $R_{\rm rms}$ & \multicolumn{1}{c}{${\cal O}_J(\beta_\perp, \beta_z)$} \\
\hline
 & \raisebox{-1.8ex}[0pt][0pt]{ND} & \multicolumn{2}{c}{$-61.00$ $(1.5, 2.8)$} & \multicolumn{1}{c}{$6.55$} & $2.31$ & $2.57$ & $2.33$ & \multicolumn{2}{c}{$0.995$ $(1.6, 3.0)$} \\
\raisebox{-1.8ex}[0pt][0pt]{$0^+$} &  & \multicolumn{2}{c}{$-61.00$ $(2.3, 1.1)$} & \multicolumn{1}{c}{$6.55$} & $2.31$ & $2.57$ & $2.33$ & \multicolumn{2}{c}{$0.995$ $(2.4, 0.9)$} \\
 & \raisebox{-1.8ex}[0pt][0pt]{JA} & \multicolumn{2}{c}{$-59.62$ $(1.6, 2.8)$} & \multicolumn{1}{c}{$5.17$} & $2.33$ & $2.72$ & $2.36$ & \multicolumn{2}{c}{$0.993$ $(1.5, 3.1)$} \\
 &  & \multicolumn{2}{c}{$-59.62$ $(2.4, 1.0)$} & \multicolumn{1}{c}{$5.17$} & $2.33$ & $2.72$ & $2.36$ & \multicolumn{2}{c}{$0.993$ $(2.5, 0.5)$} \\
\hline
 & \raisebox{-1.8ex}[0pt][0pt]{ND} & \multicolumn{2}{c}{$-58.10$ $(0.3, 2.8)$} & \multicolumn{1}{c}{$3.65$} & $2.29$ & $2.55$ & $2.31$ & \multicolumn{2}{c}{$0.994$ $(0.1, 3.0)$} \\
\raisebox{-1.8ex}[0pt][0pt]{$2^+$} &  & \multicolumn{2}{c}{$-58.06$ $(2.2, 0.2)$} &\multicolumn{1}{c}{$3.61$} & $2.30$ & $2.57$ & $2.32$ & \multicolumn{2}{c}{$0.991$ $(2.3, 0.2)$} \\
 & \raisebox{-1.8ex}[0pt][0pt]{JA} & \multicolumn{2}{c}{$-56.68$ $(0.3, 2.9)$} & \multicolumn{1}{c}{$2.23$} & $2.31$ & $2.70$ & $2.34$ & \multicolumn{2}{c}{$0.991$ $(0.1, 3.2)$} \\
 &  & \multicolumn{2}{c}{$-56.64$ $(2.2, 0.2)$} & \multicolumn{1}{c}{$2.19$} & $2.30$ & $2.69$ & $2.33$ & \multicolumn{2}{c}{$0.987$ $(2.4, 0.2)$} \\
\hline
 & \raisebox{-1.8ex}[0pt][0pt]{ND} & \multicolumn{2}{c}{$-50.47$ $(0.1, 2.1)$} & \multicolumn{1}{c}{$-3.98$} & $2.23$ & $2.51$ & $2.24$ & \multicolumn{2}{c}{$0.977$ $(0.1, 2.1)$} \\
\raisebox{-1.8ex}[0pt][0pt]{$4^+$} &  & \multicolumn{2}{c}{$-50.23$ $(1.6, 0.2)$} & \multicolumn{1}{c}{$-4.22$} & $2.20$ & $2.49$ & $2.22$ & \multicolumn{2}{c}{$0.967$ $(1.7, 0.2)$} \\
 & \raisebox{-1.8ex}[0pt][0pt]{JA} & \multicolumn{2}{c}{$-48.96$ $(0.1, 2.1)$} & \multicolumn{1}{c}{$-5.49$} & $2.23$ & $2.63$ & $2.26$ & \multicolumn{2}{c}{$0.974$ $(0.1, 2.2)$} \\
 &  & \multicolumn{2}{c}{$-48.70$ $(1.6, 0.2)$} & \multicolumn{1}{c}{$-5.75$} & $2.20$ & $2.60$ & $2.23$ & \multicolumn{2}{c}{$0.963$ $(1.7, 0.2)$} \\
\hline\hline
\end{tabular}
\end{center}
\end{table}

In order to compare the single Hyper-THSR wave function with the Brink-GCM wave function, we calculate the following squared overlap:
\begin{equation}
{\cal O}_J (\beta_\perp, \beta_z) = \frac{|\langle \Psi^H_J(\beta_\perp, \beta_z) | {\widetilde \Psi}^B_J \rangle|^2}{\langle \Psi^H_J(\beta_\perp, \beta_z) | \Psi^H_J(\beta_\perp, \beta_z) \rangle \langle {\widetilde \Psi}^B_J | {\widetilde \Psi}^B_J \rangle }, \label{eq:ovlp}
\end{equation}
with
\begin{equation}
{\widetilde \Psi}^B_J= \sum_{R,S} f^{(J,\lambda=0)}(R,S) [u_J(\vc{R}), \psi_{\lambda=0}(\vc{S})]_J.
\end{equation}
The above wave function ${\widetilde \Psi}^{B}_J$ is the Brink-GCM wave function projected onto the model space with the angular-momentum channel $(L,\lambda)=(J,0)$.
%This justifies the present model space for the Hyper-THSR wave function, where only $S$-wave component for the $\Lambda$ particle is taken into account. 
\begin{figure}[htbp]
\begin{center}
\includegraphics[scale=0.70]{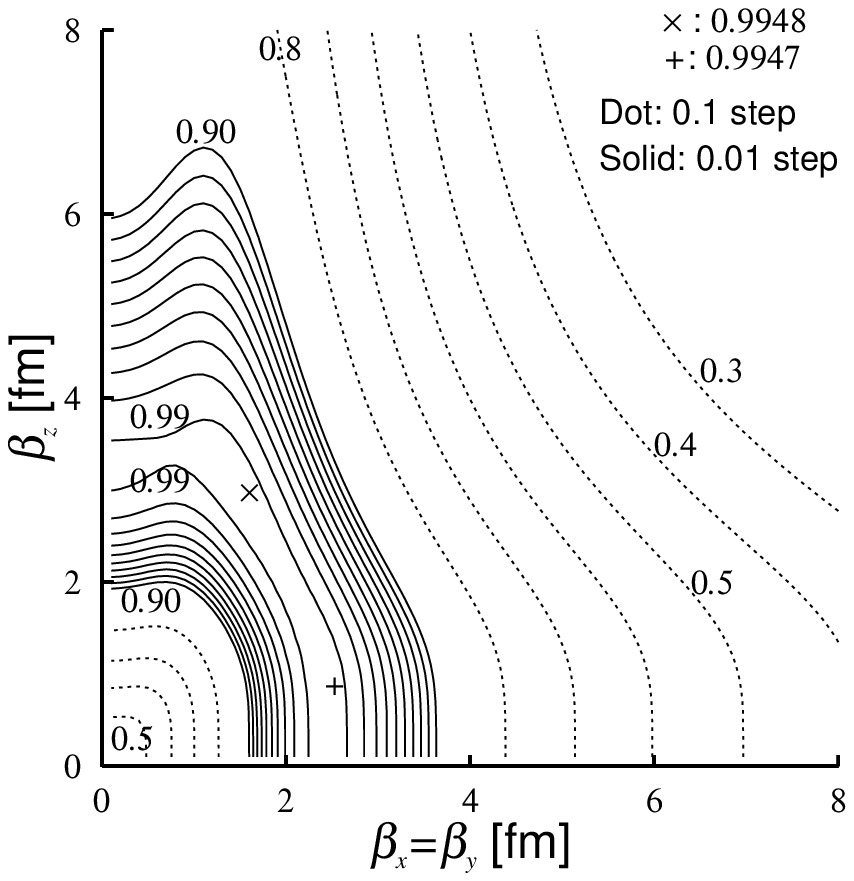}
\includegraphics[scale=0.70]{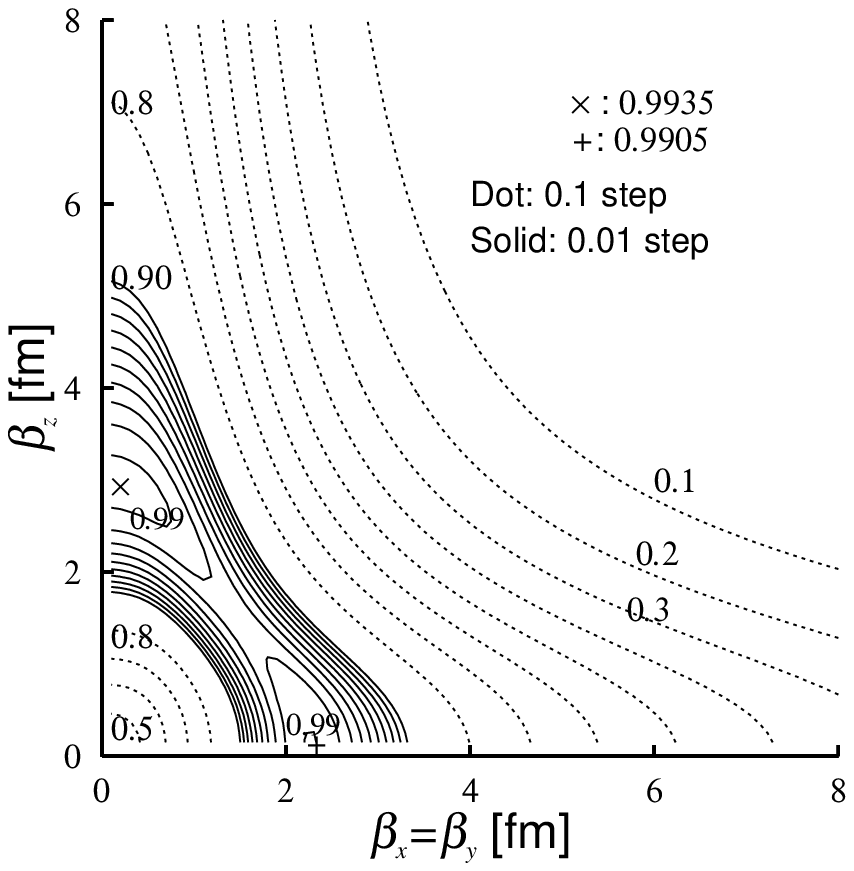}
\includegraphics[scale=0.70]{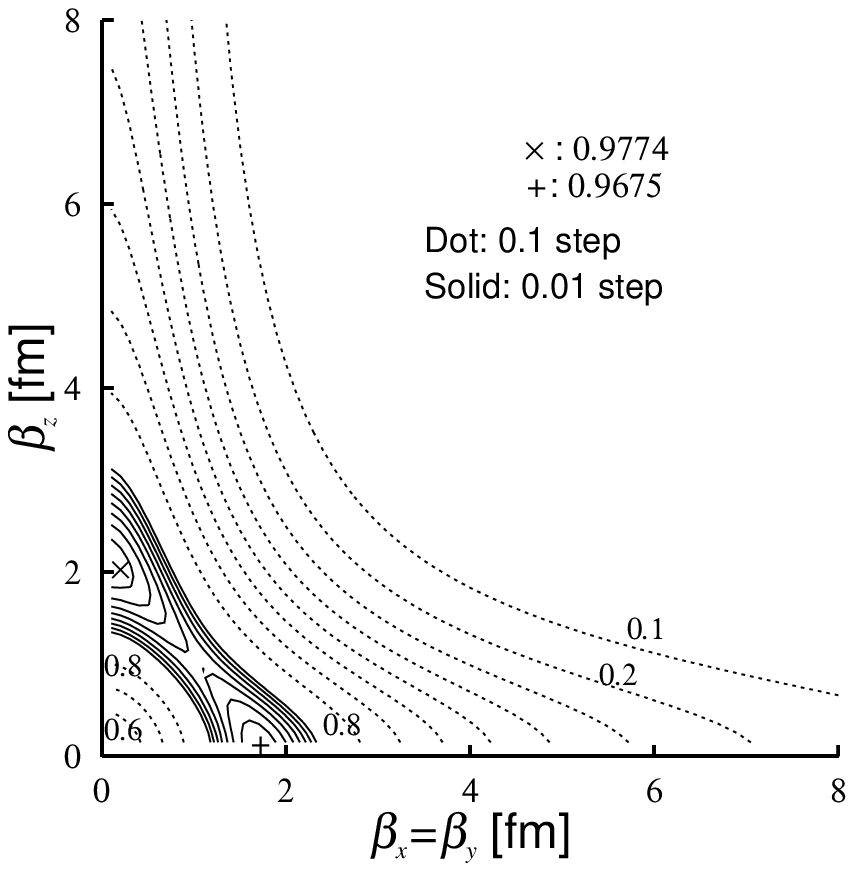}
\caption{Contour maps of the squared overlap surfaces for $J^\pi=0^+$(left top), $2^+$(right top), and $4^+$(bottom) states in two-parameter space $\beta_x=\beta_y(\equiv \beta_\perp), \beta_z$, defined by ${\cal O}(\beta_\perp,\beta_z)$ in Eq.~(\ref{eq:ovlp}). Two maxima are denoted by $\times$ and $+$. YNG-ND interaction is adopted for the $\Lambda N$ interaction.}\label{fig:ovlp}
\end{center}
\end{figure}
In Fig.~\ref{fig:ovlp}, we show the contour maps of this quantity in the two-parameter space $\beta_\perp$ and $\beta_z$, calculated for the $J^\pi=0^+$ (left top), $2^+$ (right top), and $4^+$ (bottom) states with the choice of potential ND. For these states, we can see two maxima in prolately deformed and oblately deformed regions (denoted as $\times$ and $+$), as in the case of the energy surfaces. The values of the maxima are extremely large and close to unity. The values for the $J^\pi=0^+,2^+$ states are $99.5$ \% and $99.4$ \%, respectively. For the $J^\pi=4^+$ state, the maximum value is slightly down to $97.7$ \%. This reduction may originate from a mixture of spurious scattering-state components. The maximum values and $\vc{\beta}$ values giving them are listed in Table~\ref{tab2}. We can see that the $\vc{\beta}$ values giving the minimum energies and the maximum squared overlaps almost coincide with each other. These practically $100$ \% squared overlaps of course mean that, at least on the subspace with $L=J$ and $\lambda=0$, the Brink-GCM wave function obtained by solving the $2\alpha + \Lambda$ Hill-Wheeler equation is equivalent to the single configuration of the Hyper-THSR wave function. We should note that the Brink-GCM wave function in this subspace $(J,\lambda)=(L,0)$ is shown to be almost the same as the one in the full angular-momentum-channel space (see Table~\ref{tab1}). This result also means that the size parameter $\vc{\beta}$, which specifies the monopole-like dilatation of the whole nucleus, quite well takes into account the effect of the spatial core shrinkage by the additional $\Lambda$ particle. We can thus conclude that in ${^9_\Lambda {\rm Be}}$ the $2\alpha$ clusters are trapped into a container, which is specified by the optimal value of the size parameter $\vc{\beta}$, under the influence of the antisymmetrizer ${\cal A}$ acting on the nucleons, such as realized in the form of the wave function Eq.~(\ref{eq:hypthsr}).  

Next a question arises how this container picture is justified for the $\Lambda$ particle. Since now we know that the ${^8{\rm Be}}$ core can be described by only the single configuration of Hyper-THSR wave function with optimal $\vc{\beta}$ value in Eq.~(\ref{eq:hypthsr}), the $\Lambda$-particle wave function can be described separately from the core part, as $\sum_\kappa f_\Lambda (\vc{\beta},\kappa)\varphi(\kappa)$ in Eq.~(\ref{eq:hypthsr}). This means that the total wave function retains the product nature for the $\alpha$$\alpha$$\Lambda$ clusters, and therefore the container structure is kept for the $\Lambda$ particle as well as the $2\alpha$ clusters. The $\Lambda$-particle wave function must be an eigenfunction of $\Lambda$-${^8{\rm Be}}$ folding potential. We should also note that the $\Lambda$-particle wave functions for $J^\pi=0^+,2^+,4^+$ states are still practically identical to the single Gaussian functions. We calculate the maximum values for the squared overlaps with the single Gaussian functions $|\langle \sum_\kappa f_\Lambda (\vc{\beta},\kappa)\varphi(\kappa) | \varphi(\kappa_0)\rangle|^2$, which are $0.992$, $0.992$, and $0.992$, with $\kappa_0^{-1/2}=2.95$ fm, $2.93$ fm, and $2.86$ fm, for $J^\pi=0^+$, $2^+$, and $4^+$ states, respectively.

\subsection{Discussion of intrinsic structure}

We here discuss the intrinsic structure of ${^{9}_\Lambda {\rm Be}}$, together with that of ${^8{\rm Be}}$. Using the intrinsic wave function defined as the THSR wave function before the angular-momentum projection in Eq.~(\ref{eq:hypthsr}), we can calculate the following intrinsic density of nucleons,
\begin{equation}
\rho_N(\vc{r})=\frac{\big\langle \Phi_{2\alpha-\Lambda}^{\rm H-THSR}(\beta_\perp,\beta_z) \big| \sum_{i=1}^{8}\delta( \vc{r}_i-\vc{X}_G - \vc{r}) \big| 
\Phi_{2\alpha-\Lambda}^{\rm H-THSR}(\beta_\perp,\beta_z) \big\rangle}{\big\langle \Phi_{2\alpha-\Lambda}^{\rm H-THSR}(\beta_\perp,\beta_z) \big| \Phi_{2\alpha-\Lambda}^{\rm H-THSR}(\beta_\perp,\beta_z) \big\rangle}. \label{eq:dsty}
\end{equation}
This density is normalized as usual to the total number of nucleons, $\int d^3r\rho_N({\vc r}) = 8$. 
\begin{figure}[htbp]
\begin{center}
\includegraphics[scale=0.50, angle=270]{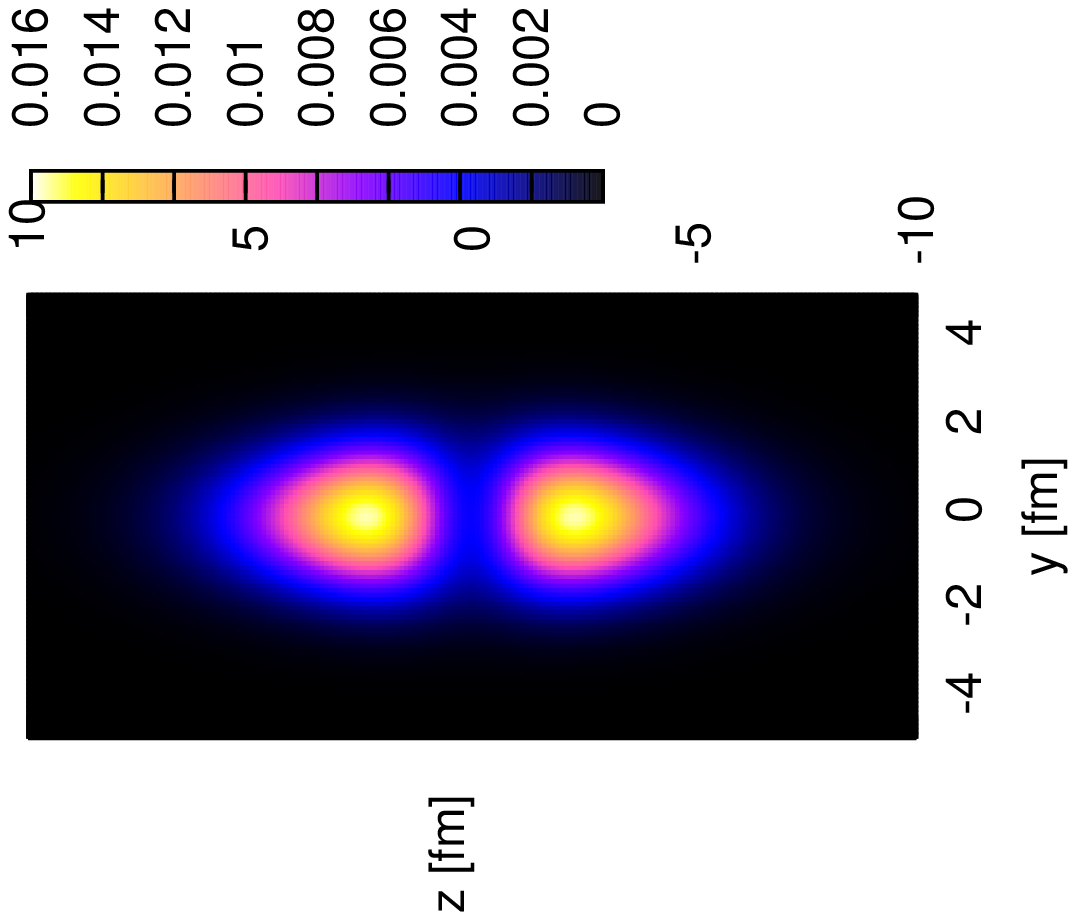}
\includegraphics[scale=0.50, angle=270]{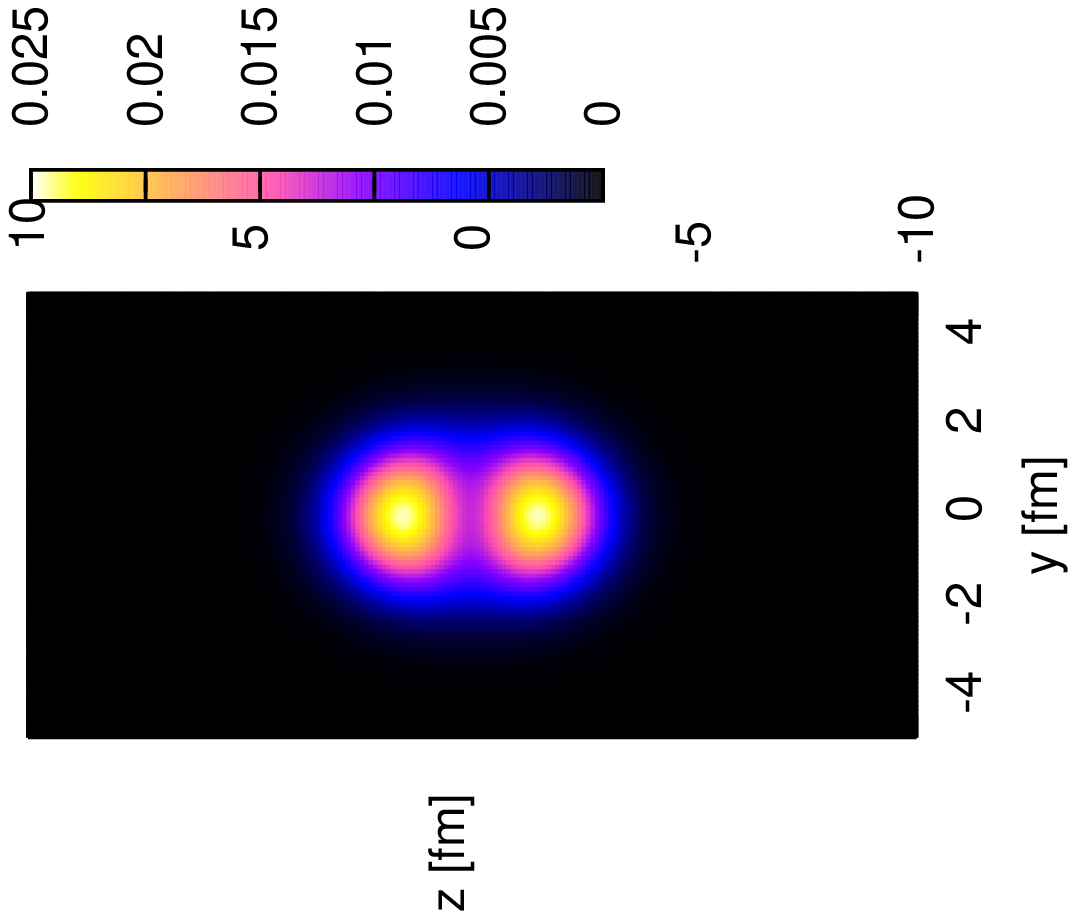}
\caption{Intrinsic density profiles of ${^9_\Lambda{\rm Be}}$ (right) defined by Eq.~(\ref{eq:dsty}), on $yz$ plane with $x=0$. For comparison, the one of ${^8{\rm Be}}$ is also shown (left).}\label{fig:dsty_9LBe} 
\end{center}
\end{figure}

In Fig.~\ref{fig:dsty_9LBe}, we show the intrinsic density profiles of ${^9_\Lambda{\rm Be}}$ at right defined by Eq.~(\ref{eq:dsty}) and of ${^8{\rm Be}}$ at left. The single optimal $\vc{\beta}$ values giving the minimum energies are adopted, i.e. $(\beta_\perp,\beta_z)=(1.5, 2.8)$ for ${^9_\Lambda{\rm Be}}$ and $(1.8, 7.8)$ for ${^8{\rm Be}}$. While both ones clearly show the $2\alpha$-cluster structure with the prolately deformed shape, ${^8{\rm Be}}$ has a gaslike tail of the $2\alpha$ clusters and ${^9_\Lambda{\rm Be}}$ not. In ${^9_\Lambda{\rm Be}}$ the $\Lambda$ particle gives rise to strong shrinkage and the gaslike tail in ${^8{\rm Be}}$ disappears. The rms radius of the ${^8{\rm Be}}$ core is accordingly changed from $R_{\rm rms}=2.87$ fm for ${^8{\rm Be}}$ to $R^{\rm (c)}_{\rm rms}=2.31$ fm for ${^9_\Lambda{\rm Be}}$. Nevertheless the $2\alpha$-cluster structure definitely remains in this very compact object, which is produced by the competition between the quite strong effect of inter-$\alpha$ Pauli repulsion originating from the antisymmetrizer ${\cal A}$ and the fairly strong attractive effect among the $2\alpha$ and $\Lambda$ particles, and therefore the $2\alpha$ clusters in this intrinsic state are effectively localized in space. 
This means that even for the states which are described by nonlocalized-type wave function with container structure, localized nature of clustering can appear in density distribution due to the Pauli principle. We can say that dynamics prefers nonlocalized clustering but kinematics coming from the Pauli principle makes the system look like localized clustering in ${^{9}_\Lambda{\rm Be}}$.

\begin{figure}[htbp]
\begin{center}
\includegraphics[scale=0.50, angle=270]{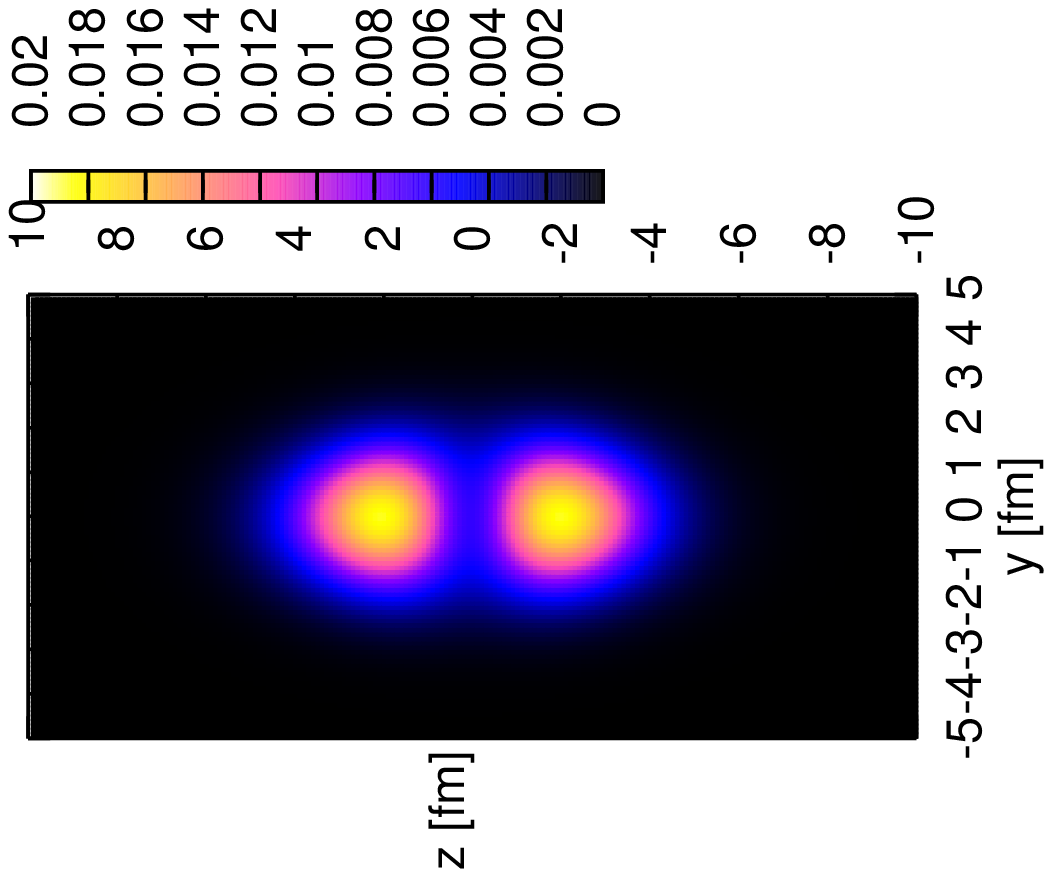}
\includegraphics[scale=0.50, angle=270]{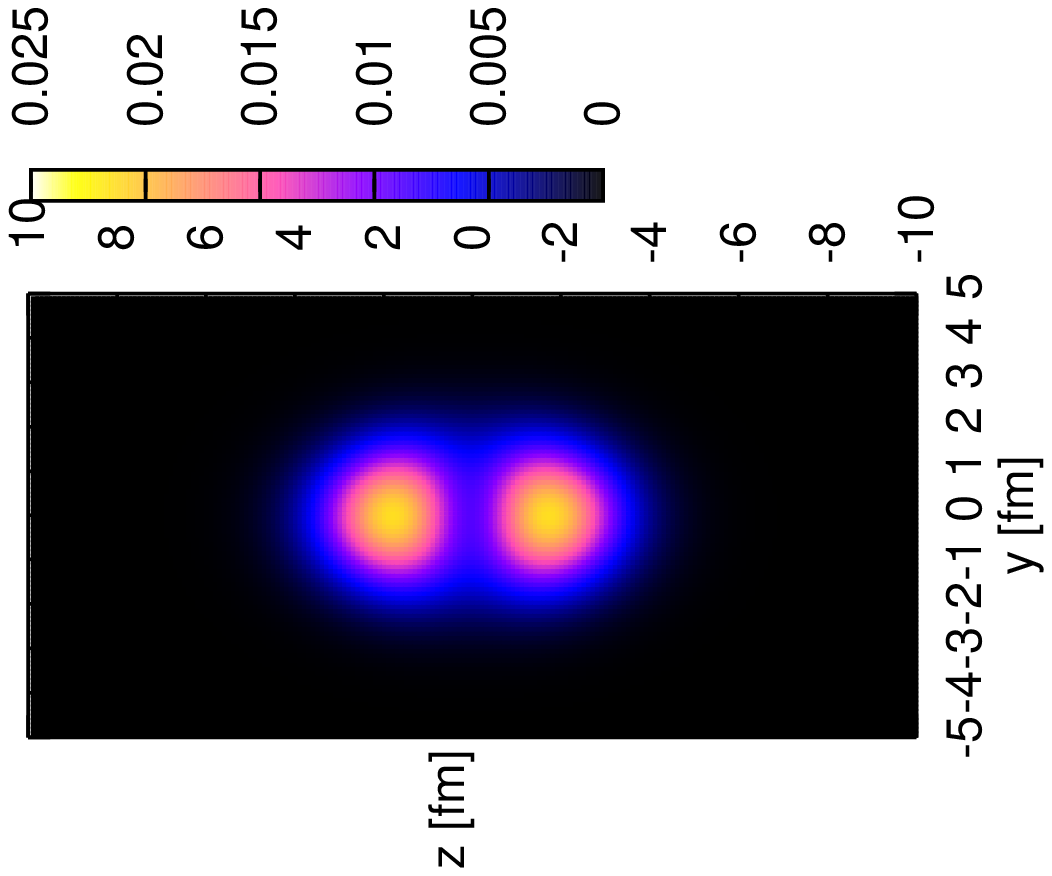}
\includegraphics[scale=0.50, angle=270]{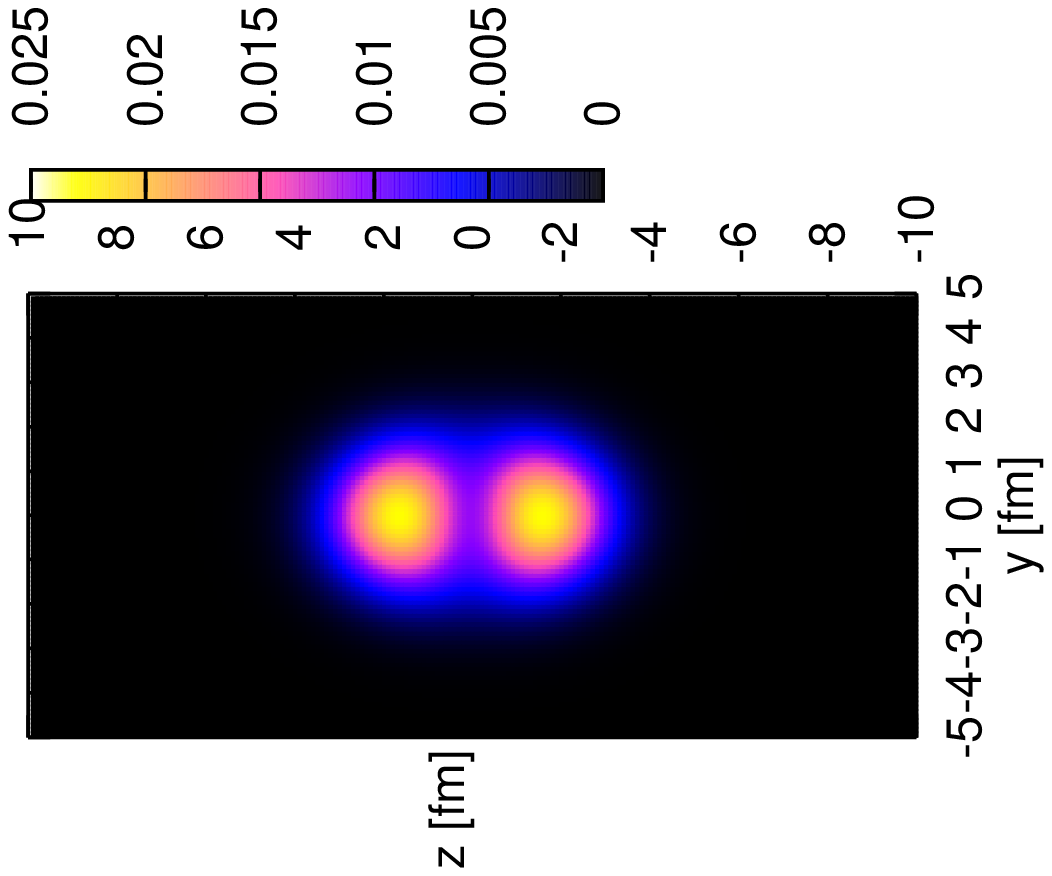}
\caption{Intrinsic density profiles of ${^9_\Lambda{\rm Be}}$ defined by Eq.~(\ref{eq:dsty}), on $yz$ plane with $x=0$, with the uses of artificial $\Lambda N$ interaction, $V_i^{(\Lambda N)} \rightarrow \delta \times V_i^{(\Lambda N)}$ in Eq.~(\ref{eq:hml}), for $\delta=$$0.4$ (left), $0.6$ (middle), and $0.8$ (right). Potential ND is used for $V_i^{(\Lambda N)}$.}\label{fig:dsty1_9LBe} 
\end{center}
\end{figure}

\begin{table}[htbp]
\begin{center}
\caption{The minimum binding energies of $E(\beta_\perp,\beta_z)$, the corresponding $\Lambda$ binding energies $B_\Lambda$, the rms radii of the ${^8{\rm Be}}$ core $R^{\rm (c)}_{\rm rms}$, the rms distances between the core and $\Lambda$ particle $R^{({\rm c}-\Lambda)}_{\rm rms}$, and the rms radii of ${^9_\Lambda{\rm Be}}$, $R_{\rm rms}$, at the minimum positions, calculated by using artificial $\Lambda N$ interaction $V_i^{(\Lambda N)} \rightarrow \delta \times V_i^{(\Lambda N)}$ in Eq.~(\ref{eq:hml}). The maximum squared overlap values of ${\cal O}_J(\beta_\perp,\beta_z)$ defined by Eq.~(\ref{eq:ovlp}), are also shown. Potential ND is used for $V_i^{(\Lambda N)}$. $\delta=0$ corresponds to the results of ${^{8}{\rm Be}}$. \label{tab3}}
\begin{tabular}{cclcccccl}
\hline\hline
$\delta$ & \multicolumn{2}{c}{$E(\beta_\perp,\beta_z)$} & \multicolumn{1}{c}{$B_\Lambda$} & $R^{\rm (c)}_{\rm rms}$ & $R^{{\rm (c}-\Lambda)}_{\rm rms}$ & $R_{\rm rms}$ & \multicolumn{2}{c}{${\cal O}_{J=0}(\beta_\perp,\beta_z)$} \\
\hline
$0.0$ & \multicolumn{2}{c}{$-54.45$$(1.8,7.8)$} &  & $2.87$ &  &  & \multicolumn{2}{c}{$1.00$ $(1.8,7.8)$} \\
$0.4$ & \multicolumn{2}{c}{$-54.67$$(1.8,5.4)$} & \multicolumn{1}{c}{$0.22$} & $2.65$ & $7.13$ & $3.36$ & \multicolumn{2}{c}{$0.994$$(1.7,5.9)$} \\
$0.6$ & \multicolumn{2}{c}{$-56.09$$(1.7,3.9)$} & \multicolumn{1}{c}{$1.64$} & $2.47$ & $3.78$ & $2.62$ & \multicolumn{2}{c}{$0.993$$(1.7,4.2)$} \\
$0.8$ & \multicolumn{2}{c}{$-58.29$$(1.6,3.2)$} & \multicolumn{1}{c}{$3.84$} & $2.38$ & $2.95$ & $2.43$ & \multicolumn{2}{c}{$0.994$$(1.6,3.4)$} \\
$1.0$ & \multicolumn{2}{c}{$-61.00$$(1.5,2.8)$} & \multicolumn{1}{c}{$6.55$} & $2.31$ & $2.57$ & $2.33$ & \multicolumn{2}{c}{$0.995$$(1.6,3.0)$} \\
\hline
\end{tabular}
\end{center}
\end{table}

One of the reasons why this drastic shrinkage happens is that the $\Lambda$ particle is out of the antisymmetrization and can stay deeply inside the core nucleus to gain deeper binding energy.  
We then simulate the shrinkage effect by varying $\Lambda N$ interaction artificially, with overall factor $\delta$ multiplied with $V_i^{(\Lambda N)}$ in Eq.~(\ref{eq:hml}). In Fig.~\ref{fig:dsty1_9LBe}, the intrinsic density profiles calculated for $\delta=0.4$ (left), $0.6$ (middle), and $0.8$ (right) are shown. The $\vc{\beta}$ values to give the energy minima after projection onto $J=0$ space, which are listed in Table~\ref{tab3}, are adopted. As the $\Lambda N$ interaction is strengthened, from $\delta=0.4$ to $0.8$, the $\alpha$-$\alpha$ distance is shortened along $z$-direction. The corresponding $\vc{\beta}$ values, $\Lambda$ binding energies, the maximum squared overlaps in Eq.~(\ref{eq:ovlp}) with the use of the same artificial $\Lambda N$ potential, rms radii of the ${^{8}{\rm Be}}$ core, etc, like in Table~\ref{tab2}, are also shown in Table~\ref{tab3}. $\delta=0$ and $1$ correspond to the cases of ${^8{\rm Be}}$ and ${^{9}_\Lambda {\rm Be}}$, respectively. Starting from the case of ${^8{\rm Be}}$ with very large value of $\beta_z$ and small value of $\beta_\perp$, i.e. $\beta_\perp=1.8$ fm and $\beta_z=7.8$ fm, only the $\beta_z$ value drastically gets smaller as the increase of $\Lambda N$ interaction, and eventually for ${^{9}_\Lambda {\rm Be}}$, $\beta_z$ becomes much smaller while $\beta_\perp$ is almost unchanged, i.e. $\beta_\perp=1.5$ fm and $\beta_z=2.8$ fm. 
In the case of $\delta=0.4$ the $\Lambda$ binding energy is very small $B_\Lambda=0.22$ MeV and accordingly the rms distance of the $\Lambda$ particle from the core is very long, $R^{({\rm c}-\Lambda)}_{\rm rms}=7.13$ fm, since the $\Lambda$ particle cannot be trapped into the Coulomb barrier. As the increase of the $\Lambda$ binding energy, the distance $R^{({\rm c}-\Lambda)}_{\rm rms}$ gets shorter and the $\Lambda$ particle gets to localize around the core to gain the energy. At $\delta=1.0$, where the $\Lambda N$ interaction is strong enough, since the $\Lambda$ particle is free from the antisymmetrizer, it can come in deeply inside the core, so that it causes the strong shrinkage of the core and the very compact object is realized. Nevertheless, due to the strong effect of the antisymmetrization between nucleons, the core keeps very clear spatial localization of the $2\alpha$ clusters.
We should note that at each step of the change of $\Lambda N$ interaction, from dilute $2\alpha$ cluster structure to compact localized $2\alpha$ cluster structure, all are described precisely by single Hyper-THSR wave functions, which have optimal sizes of deformed container, i.e. those with large-size container to small-size container. We should emphasize that at every step of $\delta$, the squared overlaps with the full Brink-GCM solutions are more than $99.3$\%. This may imply that the THSR-type container picture is essentially important in understanding every type of cluster structures, from the compact to dilute ones.

As mentioned in \S~\ref{intro}, the THSR wave function does not only describe accurately the gaslike cluster states but also non-gaslike cluster states such as the inversion doublet bands of $\alpha + {^{16}{\rm O}}$ in ${^{20}{\rm Ne}}$, $3\alpha$ and $4\alpha$ linear-chain states, etc. These states commonly retain the above mentioned nature of the container structure with the effective localization coming from the Pauli principle. This is considered to be a key ingredient to understand nuclear cluster structures. The ${^9_\Lambda {\rm Be}}$ nucleus is one of the most compact objects with cluster structure and hence is quite different from ${^8{\rm Be}}$ with dilute $2\alpha$-cluster structure. 
Thus the present result indicates that this picture is also indispensable to understand hypernuclear cluster structures, even in a high density limit of clustering, where the effect of the Pauli principle on nucleons is too strong, due to the shrinkage effect of the $\Lambda$ particle which is free from the Pauli principle and hence is not prevented from sitting deeply inside the core.
%Thus the present result that the ${^9_\Lambda {\rm Be}}$ can also be described very precisely by the single Hyper-THSR wave function supports our idea that the picture of the dynamically nonlocalized clustering, together with kinematically effective localization, is also essentially important in nuclear clustering phenomena with hyperon and under a kind of high density limit of clustering, which are very strongly influenced by the Pauli principle. 
This further implies that this concept might also play an important role in describing cluster structures in neutron-rich nuclei, since additional neutrons are also expected to play a glue-like role in core nuclei, although the neutrons are antisymmetrized with nucleons in core nuclei.

\section{Summary}\label{summary}
We newly introduced the Hyper-THSR wave function, which can be applied to hypernuclei and takes over the important feature of the original THSR wave function that the constituent clusters are confined into a container, whose size is the variational parameter $\vc{\beta}$, under the consideration of the effect of the antisymmetrization of nucleons. 
%This is natural extension of the THSR wave function which has succeeded in describing cluster states in ordinary nuclei to the systems with additional $\Lambda$ particle. 
We investigated ${^9_\Lambda {\rm Be}}$ using this new cluster model wave function. We first performed the full $2\alpha + \Lambda$ Brink-GCM calculation and compared the solutions for $J^\pi=0^+$, $2^+$, and $4^+$ states with the corresponding Hyper-THSR wave functions. The dominant components of the Brink-GCM wave function are shown to be the ones from $(L,\lambda)=(J,0)$ channel, where the $\Lambda$ particle couples to the core in an $S$-wave. We showed that the components of the $S$-wave channel of the $\Lambda$ particle are almost equivalent to the single configuration of the Hyper-THSR wave function. The squared overlaps between them are $99.5$ \%, $99.4$ \%, and $97.7$ \% for $J^\pi=0^+$, $2^+$, and $4^+$ states, respectively, with the use of potential ND. We then discussed the intrinsic structure of the $2\alpha$ part in ${^9_\Lambda {\rm Be}}$ by using the intrinsic Hyper-THSR wave function before the angular-momentum projection. The structural change by adding the $\Lambda$ particle is particularly demonstrated by artificially varying the $\Lambda N$ interaction. As the increase of the $\Lambda N$ interaction, the $\Lambda$ particle, which is free from the antisymmetrizer, gets to stay more deeply inside the core to gain larger $\Lambda$ binding energy, and causes the strong shrinkage of the core and makes it very compact object. However, the inter-cluster Pauli repulsion, due to the antisymmetrization between the nucleons, is so strong that the clear $2\alpha$-cluster structure still survives. During the structural change of the core by varying the $\Lambda N$ interaction, the full solutions obtained by the Brink-GCM calculations can be expressed by the single Hyper-THSR wave functions, with more than $99.3$ \% accuracy. These results mean that the above mentioned container structure inherent in the Hyper-THSR wave function is exactly realized in ${^9_\Lambda{\rm Be}}$, which is made very much compact by the additional $\Lambda$ particle, and hence not at all ``gaslike'' object like ${^{8}{\rm Be}}$. These also indicate that the effect of spatial core shrinkage, invoked by the additional $\Lambda$ particle, is very nicely taken into account in the Hyper-THSR wave function, in which the size parameter $\vc{\beta}$ specifies the monopole-like dilatation of the whole nucleus. Thus this Hyper-THSR ansatz is very promising in studying cluster structures in hypernuclei. The application to heavier hypernuclei like ${^{13}_\Lambda{\rm C}}$ will be reported in a forthcoming paper. The present Hyper-THSR ansatz is very flexible. One way to push forward this container picture with inclusion of the angular-momentum channels other than the $S$-wave is to consider the deformation for the $\Lambda$-particle wave function as well. The extension to a coupled-channel approach is also not difficult, and they are left as future works. 

\section*{Acknowledgment}
The authors wish to thank H. Horiuchi, A. Tohsaki, P. Schuck, G. R\"opke, Z. Z. Ren, and C. Xu for many helpful discussions concerning this work. This work was suported by JSPS KAKENHI Grants No.~23224006 and No.~25400288, RIKEN iTHES Project, and HPCI Strategic Program of Japanese MEXT.


%\bibliographystyle{ptephy}
%\bibliography{sample}
%
% once the .bbl file has been generated then place the text in your article.

\vfill\pagebreak

\end{document}